\documentclass[iop,letter]{emulateapj}
\usepackage[flushleft]{threeparttable}
\usepackage{subfigure}
\usepackage{graphicx}
\shorttitle{CHVC near the Magellanic Stream}
\usepackage[utf8]{inputenc}
\shortauthors{Kumari et al. }
\usepackage{hyperref}
\hypersetup{colorlinks=false}

\begin{document}

\title{A Compact High Velocity Cloud near the Magellanic Stream: Metallicity and Small-Scale Structure\footnotemark[1]}
\footnotetext[1]{Based on observations taken under program 12204 of the NASA/ESA Hubble Space Telescope, obtained at the Space Telescope Science Institute, which is operated by the Association of  Universities for Research in Astronomy, Inc., under NASA contract NAS 5-26555.}

\author{Nimisha Kumari\altaffilmark{2,3}}
\author{Andrew J. Fox\altaffilmark{3}}
\author{Jason Tumlinson\altaffilmark{3}}
\author{Christopher Thom\altaffilmark{3}}
\author{Tobias Westmeier\altaffilmark{4}}
\author{Justin Ely\altaffilmark{3}}
\affil{\altaffilmark{2}Ecole Polytechnique,  Route de Saclay, 91128 Palaiseau, France}
\affil{\altaffilmark{3}Space Telescope Science Institute, Baltimore, MD 21218}
\affil{\altaffilmark{4}ICRAR, The University of Western Australia, 35 Stirling Highway, Crawley WA 6009, Australia}

\begin{abstract}
The Magellanic Stream (MS) is a well-resolved gaseous tail originating from the Magellanic Clouds. Studies of its physical properties and chemical composition are needed to understand its role in Galactic evolution. We investigate the properties of a compact HVC (CHVC 224.0-83.4-197) lying close on the sky to the MS to determine whether it is physically connected to the Stream and to examine its internal structure. Our study is based on analysis of HST/COS spectra of three QSOs (Ton S210, B0120-28, and B0117-2837) all of which pass through this single cloud at small angular separation ($\lesssim$ 0.72$^{\circ}$), allowing us to compare physical conditions on small spatial scales. No significant variation is detected in the ionization structure from one part of the cloud to the other. Using $Cloudy$ photoionization models, toward Ton S210 we derive elemental abundances of [C/H] = $-$1.21 $\pm$ 0.11, [Si/H] = $-$1.16 $\pm$ 0.11, [Al/H] = $-$1.19 $\pm$ 0.17 and [O/H] = $-$1.12 $\pm$ 0.22, which agree within 0.09 dex. The CHVC abundances match the 0.1 solar abundances measured along the main body of the Stream. This suggests that the CHVC (and by extension the extended network of filaments to which it belongs) has an origin in the MS. It may represent a fragment that has been removed from the Stream as it interacts with the gaseous Galactic halo.
\end{abstract}

\section{Introduction} 

\indent The processes of inflow, outflow, and recycling of gas are key to a full understanding of galaxy evolution. In the halo of the Milky Way, these gas-flow processes can be traced via the gaseous high-velocity clouds (HVCs), moving at $\gtrsim $ 100 km s$^{-1}$ in the local standard of rest, spread throughout the  halo of the Milky Way. They are characterized by high radial velocities incompatible with a participation in the rotation of the Galaxy. 

\indent  One of the best studied HVCs is the Magellanic Stream, a large and well-resolved HVC, discovered in 21 cm emission \citep{WW72, M74, M77}. It has a length  of  $\approx 140^\circ$ or $\approx 220^\circ$ when including the Leading Arm \citep{Hu88, BrTh04, Bruns05, N10}, and its different regions show different phases, e.g. molecular hydrogen in the main body \citep{Rich01} and in the Leading Arm \citep{Sem01}, regions of cold neutral \citep{Mat09} and warm neutral \citep{Bruns05} gas, and other areas containing warm-ionized \citep{Lu94, Lu98, WeWi96, Pu03} and highly ionized \citep{Sem03, Fox05, Fox10} plasma. The different parts of this multi-phase extended structure  can be studied by QSO absorption-line spectroscopy. 

\indent This paper presents the study of a compact ($\sim 1$ sq. deg.) high-velocity cloud, CHVC 224.0-83.4-197, lying near the Magellanic Stream  in the vicinity of the south galactic pole, which has a chance alignment with 3 bright QSOs, Ton S210, QSO B0120-28 (also known as QSO1220-2859), and B0117-2837 (also known as MS0117-2837). The compact HVCs (by definition) are those with angular diameter of $<$ 2$^\circ$ \citep{BrBu99}. One of the targets, Ton S210 was studied using \emph{Far Ultraviolet Spectroscopic Explorer (FUSE)} observations and H I 21 cm emission observations obtained with the Parkes telescope, giving a metallicity limit of (O/H)$<$0.46 solar at a confidence of 3$\sigma$ \citep{Sem02}. Further investigation of the same CHVC using the Space Telescope Imaging Spectrograph (STIS) installed on the \emph{Hubble Space Telescope (HST)} confirmed this observation by reporting an oxygen abundance of $-$1.01 \citep{Rich09}. In this paper, we present Cosmic Origins Spectrograph (COS) data on this direction plus two additional sightlines passing through this same CHVC, which allow us to investigate its metallicity and look for small-scale variations in metal content and ionization within a single resolved halo cloud. 

\indent The paper is organized in the following way. $\S$2 describes the observations and the data reduction along with a brief description of the instruments involved. $\S$3 presents the column density measurements, chemical abundance determination and investigation of the effects of ionization using \emph{Cloudy} models. $\S$4 discusses the main results, their implication, their comparison with the previous studies and the scope of further work. $\S$5 summarizes our findings. Throughout the paper, wavelengths and velocities are presented in the kinematical local-standard-of-rest (LSR) reference frame, as defined by \citet{KeLy86}, the atomic data are taken from Morton (2003), and the solar (photospheric) abundances are taken from \citet{Asp09}. 

%--------------------OBSERVATIONS & DATA REDUCTION -------------------------------

% Booktabs require to add \usepackage{booktabs} to your document preamble
%\ref{tab QSO overview}. 
\begin{table*}[hbt]
\resizebox{\textwidth}{!}{%
\begin{threeparttable}
\caption{Summary of Sightline Properties}

\begin{tabular}{lllllccllllll}
\hline\hline
Target & \multicolumn{1}{c}{$l$}  & \multicolumn{1}{c}{$b$}  & \multicolumn{1}{c}{v$^a_{\rm min}$}     & \multicolumn{1}{c}{v$^a_{\rm max}$}    & \multicolumn{1}{l}{H I 21 cm} & \multicolumn{1}{l}{O I $\lambda$ 1302}  & \multicolumn{2}{c}{Exposure  time (s)}                                   &   & \multicolumn{2}{c}{\begin{tabular}[c]{@{}c@{}} S/N$^d$\end{tabular}}& \multicolumn{1}{c}{Detected} \\ 
\cline{8-9}\cline{11-12}
& \multicolumn{1}{c}{($^\circ$)} & \multicolumn{1}{c}{($^\circ$)} & \multicolumn{1}{c}{(km s$^{-1}$ )} & \multicolumn{1}{c}{(km s$^{-1}$)} & \multicolumn{1}{l}{Emission$^b$} & \multicolumn{1}{l}{Absorption$^c$} & \multicolumn{1}{c}{G130M} & \multicolumn{1}{c}{G160M} &  & \multicolumn{1}{c}{1393 \AA} & \multicolumn{1}{c}{1526 \AA}& \multicolumn{1}{c}{Ions} \\
 \hline
\multicolumn{1}{c}{B0117-2837} & 225.73 & $-$83.65 & $-$270 & $-$80 & $<$17.26  & $<$14.10 & 5233 & 8519 & & 22 & 16 & \begin{tabular}[c]{@{}c@{}}C II, C IV, Si II,\\  Si III, Si IV\end{tabular} \\
B0120-28 & 225.73    & $-$82.93  & $-$250   & $-$110   & 18.43 $\pm$ 0.03 & $<$14.27  & 5183 & 8519 & & 13 & 15 & \begin{tabular}[c]{@{}c@{}}C II, C IV, Si II,\\  Si III, Si IV\end{tabular} \\
Ton S210  & 224.97  & $-$83.16  & $-$280  & $-$110  & 18.36 $\pm$ 0.03 & 13.99 $\pm$ 0.19                              & 5074 & 5468 & & 31 & 22 &\begin{tabular}[c]{@{}c@{}}C II, C IV, Si II,\\  Si III, Si IV\end{tabular}
\\[1ex]

\hline
\end{tabular}
\begin{tablenotes}
\small
\item \textbf{Notes}
\item $^a$Minimum and maximum LSR velocities of CHVC absorption, used for  AOD (Apparent Optical Depth) integration.
\item $^b$log [N(H I)/cm$^{-2}$] from Parkes observations, details in Table \ref{tab H I}. 
\item $^c$log [N(O I)/cm$^{-2}$], based on the night-only data from G130M/FUVA, details in Table \ref{tab O I}. %(\emph{Appendix})
\item The uncertainty on N(H I 21 cm) is statistical. Error on N(O I 1302) includes both statistical and systematic (continuum-placement) uncertainties. Upper limits are 3$\sigma$. 
%\item $^d$List of metal lines detected in HST/COS data
\item $^d$S/N ratio per pixel.

\end{tablenotes}
\label{tab QSO}
\end{threeparttable}
}
\end{table*}

\section{Observations and Data Reduction}

Table \ref{tab QSO} gives an overview of the three sightlines under study.

\subsection{HST/COS Observations}
HST/COS observations of the three target QSOs were taken under a 14-orbit  \emph{HST} program (ID 12204, PI C. Thom), using the COS medium-resolution FUV gratings G130M/[1291,1309] and G160M/[1600,1623] settings and the Primary Science Aperture.  Their resolving power $R\approx16,000$ corresponds to a velocity resolution $\delta v\approx19$ km\,s$^{-1}$\citep{Ho14}. COS has better sensitivity compared to STIS, with a peak FUV throughput a factor of 10-30 higher. The velocity resolution is high enough to separate the CHVC absorption ($v_{LSR}$ = $-$200 km s$^{-1}$) from the strong Milky Way disk absorption at $v_{LSR} \sim $ 0 km s$^{-1}$. 

\indent The data were reduced using CALCOS version 2.12, following the procedures given in \citet{Meiring2011} and \citet{Tumlinson2011}. In brief, the reduction steps take the individual reduced and wavelength-calibrated x1d files produced by CALCOS for each exposure, align them in wavelength space using a cross-correlation analysis, and co-add them to produce a single summed spectrum for each quasar. The coaddition operates by summing the gross source and background counts in each wavelength bin in each exposure. The final spectrum is then background-subtracted, corrected for fixed pattern noise using a reference file provided by the COS instrument team, and rebinned by three pixels. This process produced reduced spectra covering the wavelength range of 1132--1798 \AA.  The S/N ratios per pixel of each spectrum measured at 1393 \AA \hspace{1pt} and at 1526 \AA  \hspace{1pt} are given in Table \ref{tab QSO}. Because of the contamination of the data resulting from the strong geocoronal emission (air-glow) in O I 1302, a second, orbital night-only reduction of the data was completed, in which the data were extracted only over those time intervals when the sun's altitude as observed by $HST$ was less than 20$^{\circ}$. Table \ref{tab O I} presents the G130M/FUVA segment for each target before and after the filtering. The continuum for each of the three lines of sight was fitted locally to the spectral regions of interest, using first-order polynomials, and then the flux was normalized by the continuum.

\subsection{21 cm observations}

\indent The H I data were taken on April 15, 2011 with the central beam of the 21-cm multibeam receiver on the 64-m Parkes radio telescope, which provides a beam size of 14.4~arcmin FWHM at 1.4~GHz. We integrated for 2 hours on Ton S210 and B0120-28, and for 2.5 hours on B0117-2837. In addition, the standard calibrator position S8 was observed for the purpose of flux calibration. The correlator setup provided a total of 4096~spectral channels across a bandwidth of 8~MHz, resulting in a velocity resolution of about $0.4~\mathrm{km \, s}^{-1}$. In order to maximize the time on source, the frequency switching method was applied, resulting in a final rms noise level of approximately 8 to 10~mK. The data were reduced using the \textsc{Livedata}/\textsc{Gridzilla} data reduction pipeline.

\begin{figure}\centering
\includegraphics[width=\linewidth]{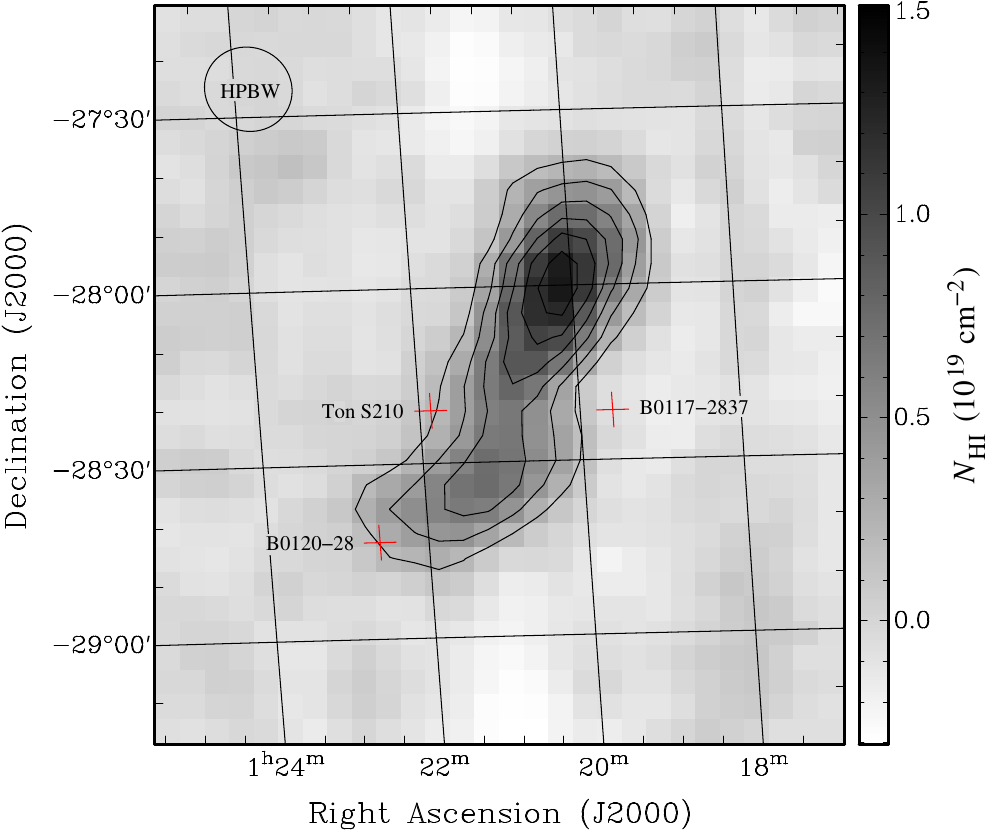}
\caption{H I 21 cm column density map of CHVC 224.0-83.4-197 from HIPASS. The positions of the three sightlines are indicated by the `$+$' signs. HPBW is the half-power beam width indicating the size of the Parkes beam, 14.4~arcmin FWHM at 1.4~GHz. The contours start at 2$\times$10$^{18}$ cm$^{-2}$ and are drawn in intervals of 2$\times$10$^{18}$ cm$^{-2}$.}
\label{fig ms cloud}
\end{figure}

  \indent Figure \ref{fig ms cloud} shows the H I 21 cm integrated map from HIPASS \citep{Bar01} centered on an LSR velocity of $-$200 km s$^{-1}$. The contours shown correspond to H I column densities of $10^{18}-10^{19}$ cm$^{-2}$. One of the three sightlines, B0117-2837 is about 7.5~arcmin from these contours. B0120-28 is farther from the center of the cloud than Ton S210, but it lies just inside the clouds's extended tail, and both of these two sightlines are nearly on the same contour. This implies the same column densities of H I for these two sightlines, and that  HIPASS data are not sensitive enough to detect emission towards B0117-2837. The original HIPASS velocity resolution is 18.0 km $s^{-1}$ (Barnes et al. 2001), while our Parkes observations have a velocity resolution of 0.4 km $s^{-1}$. Hence, there is a factor of 45 improvement in velocity resolution. In addition, we integrated for 2 hours (2.5 h in one case) as compared to the HIPASS integration time of 7.5 min (Barnes et al. 2001), which is a factor of 16 (20) in integration time, translating into an improvement in sensitivity of a factor of 4 (4.5). Figure \ref{fig channel map} presents a set of channel maps, showing the H I 21 cm emission in the region of the sky surrounding the  CHVC under study, which appears between $-$205 km s$^{-1}$ and $-$165 km s$^{-1}$.
 
\section{Results}

\begin{figure*}\centering 
%\plotone{TONS210.eps}
\includegraphics[width=0.9\linewidth]{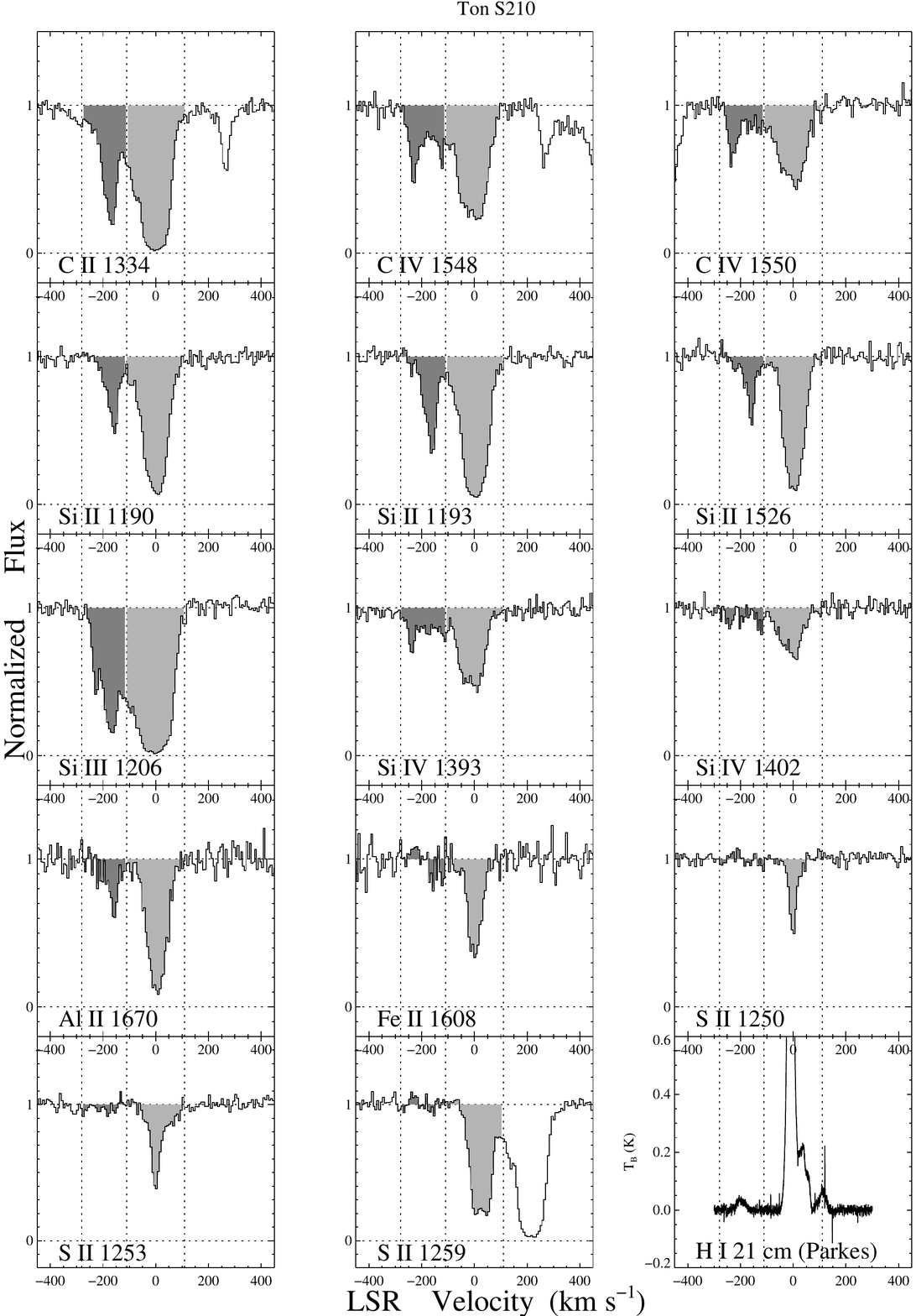}
\caption{HST/COS absorption-line profiles and Parkes H I 21 cm emission-line profile toward  Ton S210. The normalized flux is plotted against LSR velocity for each absorption line shown. The 21 cm (bottom right) panel shows the brightness temperature profile. The dark gray shaded region indicates the CHVC under study. It is enclosed by the dotted vertical lines at $-$280 km s$^{-1}$ and at $-$110 km s$^{-1}$, which are the velocity limits, $v_{min}$ and $v_{max}$ used for the AOD integration. The rightmost vertical dotted line at 110 km s$^{-1}$ marks the  limit for the Milky Way component shown in the region shaded in light gray.}
\label{fig TONS210}
\end{figure*}

\begin{figure*}\centering
%\plotone{B0120.eps} 
\includegraphics[width=0.9\linewidth]{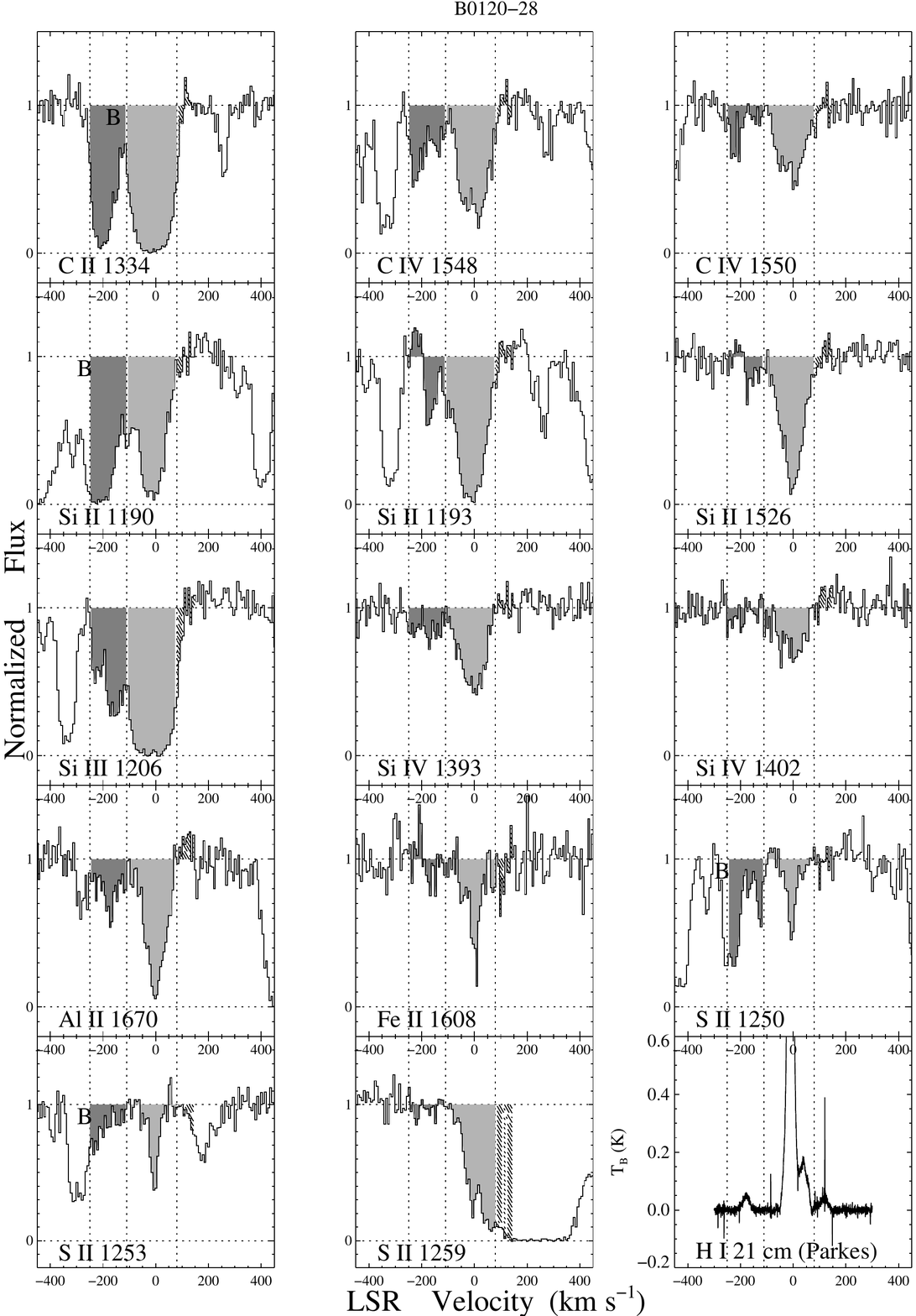}
\caption{HST/COS absorption-line profiles and Parkes H I 21 cm emission-line profile toward  B0120-28. The normalized flux is plotted against LSR velocity for each absorption line shown. The 21 cm (bottom right) panel shows the brightness temperature profile. The dark gray shaded region indicates the CHVC under study. It is enclosed by the dotted vertical lines at $-$250 km s$^{-1}$ and at $-$110 km s$^{-1}$, which are the velocity limits $v_{min}$ and $v_{max}$ used for the AOD integration. The rightmost vertical dotted line at 80 km s$^{-1}$ marks the  limit for the Milky Way component shown in the region shaded in light gray. B indicates blending.}
\label{fig B0120-28}
\end{figure*}

\begin{figure*}
\centering 
\includegraphics[width=0.9\linewidth]{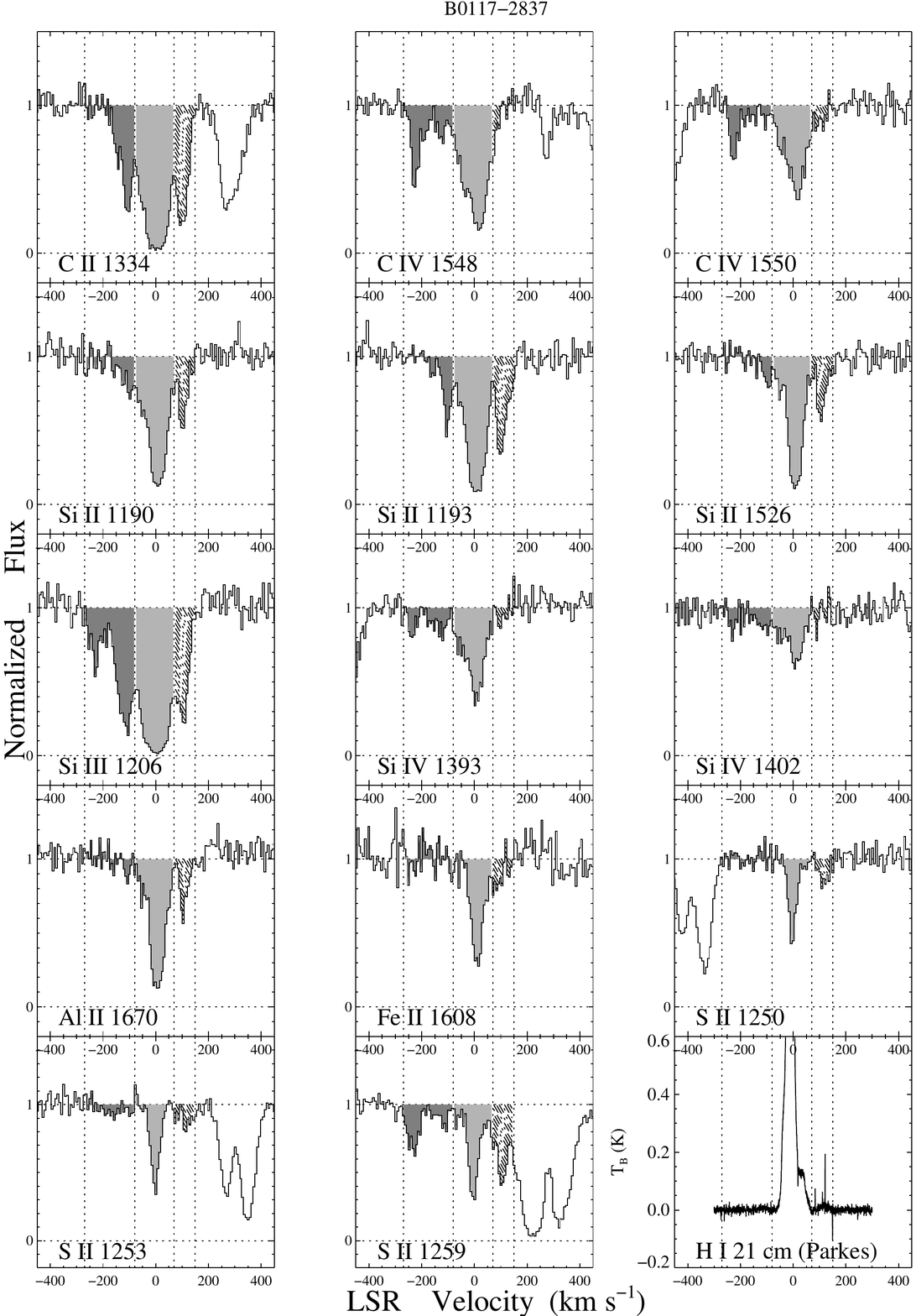}
\caption{HST/COS absorption-line profiles and Parkes H I 21 cm emission-line profile toward  B0117-2837. The normalized flux is plotted against LSR velocity for each absorption line shown. The 21 cm (bottom right) panel shows the brightness temperature profile. The dark gray shaded region indicates the CHVC under study. It is enclosed by the dotted vertical lines at $-$270 km s$^{-1}$ and at $-$80 km s$^{-1}$, which are the velocity limits $v_{min}$ and $v_{max}$ used for the AOD integration, while rightmost vertical dotted line at 70 km s$^{-1}$ marks the  limit for the Milky Way component shown in the region shaded in light gray. Another CHVC component can also be located between 70 km s$^{-1}$ and 150 km s$^{-1}$.}
\label{fig B0117-2837}
\end{figure*}

\begin{figure*}\centering 
\includegraphics[width=\linewidth]{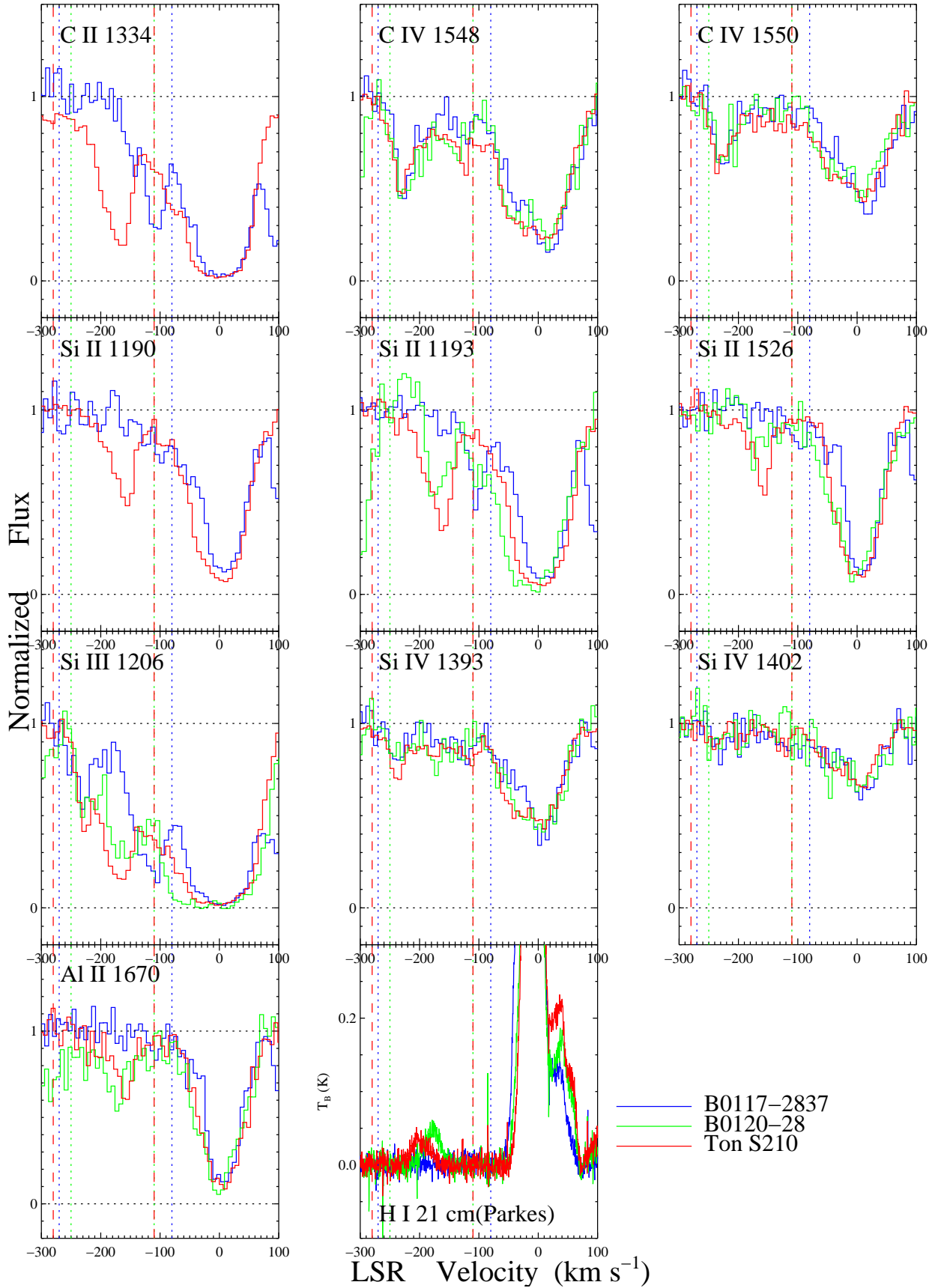}
\caption{Comparison of the absorption-line profiles of the three sightlines from HST/COS. The last panel shows the H I 21 cm brightness temperature profile from the Parkes survey. The vertical lines show velocity limits for the AOD integration, each color corresponding to a specific sightline. C II 1334 and Si II 1190 toward B0120-28 are blended and not shown on this plot. }
\label{fig comparison}
\end{figure*}

\begin{figure}[hbt]\centering 
\includegraphics[scale=0.7]{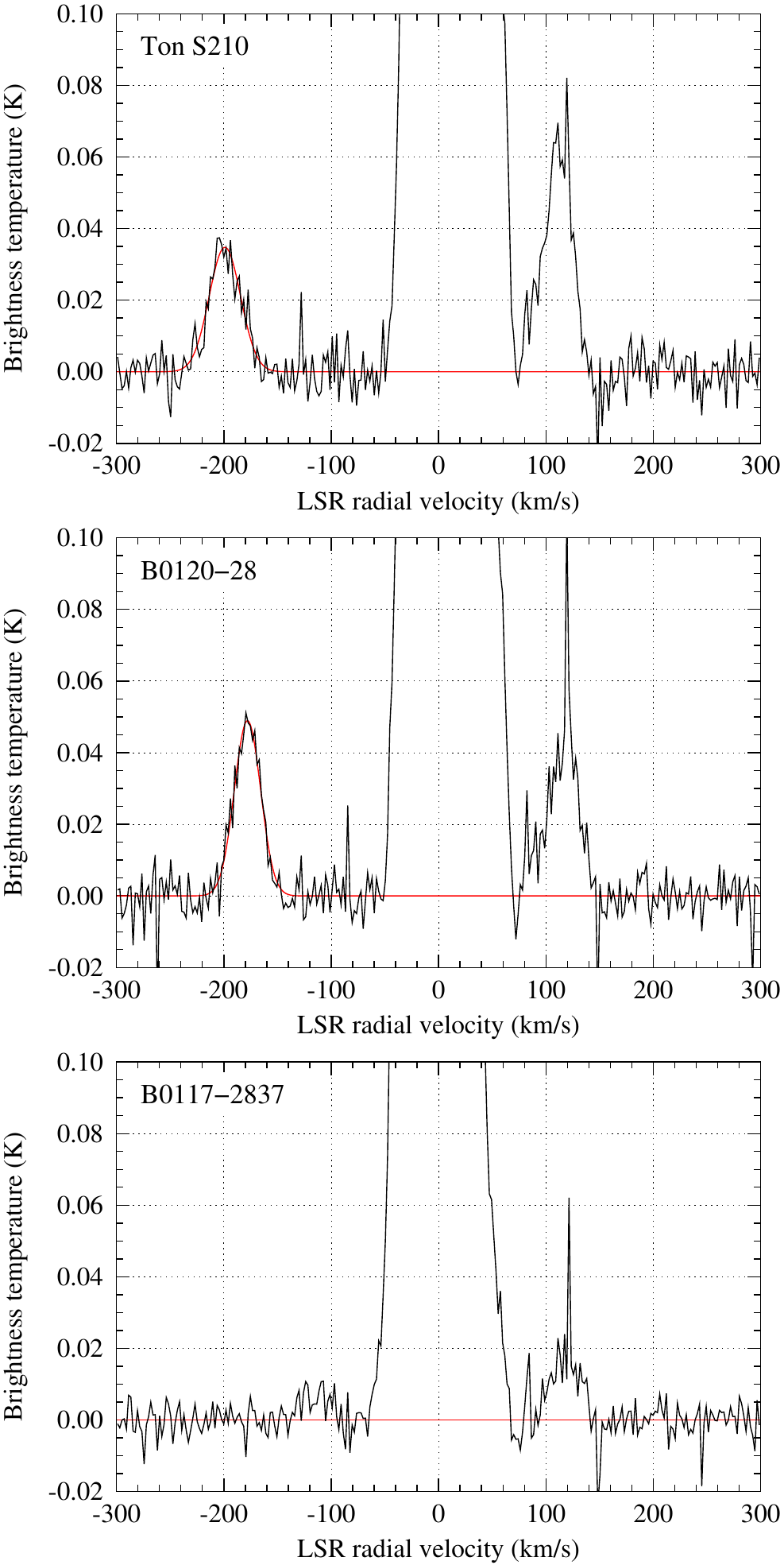}
\caption{H I 21 cm emission profiles from the Parkes radio telescope. Gaussian profile fits are shown in red, details of which can be found in Table \ref{tab H I}. Contrary to the other two sightlines, there is no H I detection for B0117-2837}
\label{fig H I}
\end{figure}

\begin{figure}[hbt]\centering 
\includegraphics[width=\linewidth]{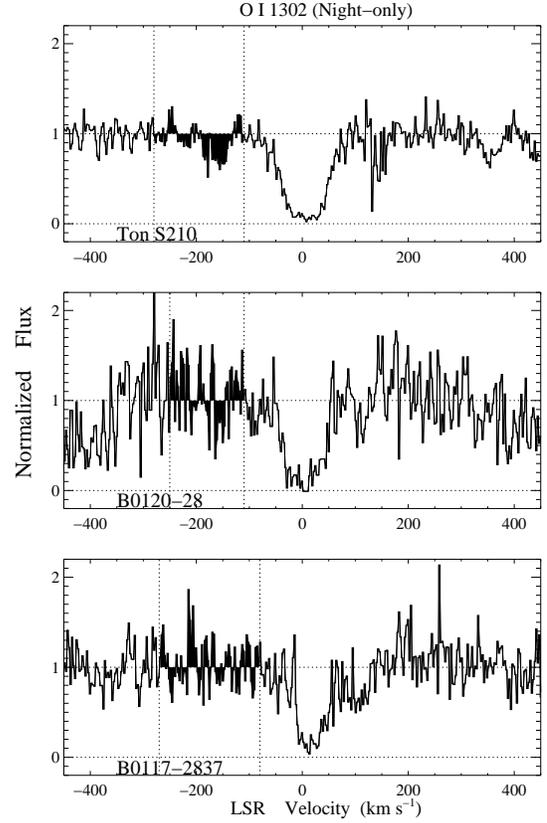}
\caption{O I 1302 absorption profile from night-only HST/COS data, details of which can be found in Table \ref{tab O I}. Of the three sightlines, only Ton S210 shows a significant detection of absorption from the CHVC (shaded in black). The dotted vertical lines are the velocity limits $v_{min}$ and $v_{max}$ used for the AOD integration.}
\label{fig nightonly}
\end{figure}

Figures  \ref{fig TONS210}, \ref{fig B0120-28}, and \ref{fig B0117-2837} show the \emph{HST/COS} data for the Ton S210, B0120-28 and B0117-2837 directions. In each case, the 21 cm emission-line profile from the Parkes radio telescope is included.  The centroid of the 21 cm velocity profile in any given direction defines where the CHVC is located in velocity in that sightline. In each plot, the absorption component corresponding to the CHVC under study is centered at a negative $v_{LSR}$ near $-$200 km s$^{-1}$, and the Milky Way component of absorption is centered at $v_{LSR}$ = 0 km s$^{-1}$. The plot of B0117-2837 shows another HVC component centered at a positive $v_{LSR}$, although there is no evidence that this component is connected to the CHVC. Therefore we do not discard this HVC, but simply do not associate it with the CHVC, since there
is no reason to do so. Since the three sightlines pass through a single small cloud, the small scale-structure variation can be studied by comparing the absorption profiles directly against one another by overplotting them in velocity space. Figure \ref{fig comparison} shows the comparison of the three sightlines with only those absorbing species which show prominent and unblended detection. C II 1334 and Si II 1190 toward B0120-28 are blended and not shown on this plot. Si II 1304 in COS data is contaminated by geocoronal emission in O I 1304 (there are O I lines at 1302 and 1304 \AA). Doing a night-only reduction removes the geocoronal emission but also reduces the S/N, meaning we can not get a good Si II column density from that line. So we do not include Si II 1304 in our study. We note that Fe II 1608, S II 1250, 1253, 1259 are not prominent toward any of the sightlines under study and hence are not shown in figure \ref{fig comparison}. We include these lines in Figures  \ref{fig TONS210}, \ref{fig B0120-28} and \ref{fig B0117-2837}, and in Table \ref{tab N measurement} for the depletion and abundance studies discussed later in $\S$4. Figure \ref{fig H I}  shows H I 21 cm emission profiles from the Parkes radio telescope. Figure \ref{fig nightonly} shows the plot of O I 1302 toward each sightline.   

\begin{table}[hbt]
\begin{threeparttable}
\caption{CHVC Column Density Measurements}
\begin{tabular}{lllll}
\hline\hline
Ion    & \begin{tabular}[c]{@{}c@{}}Line \\ (\AA)\end{tabular} & \multicolumn{3}{c}{\begin{tabular}[c]{@{}c@{}} log $\left(\frac{N^c_a}{cm^{-2}}\right)$\end{tabular}} \\ 
\cline{3-5}
       &        & Ton S210           & B0120-28             & B0117-2837    \\
\hline       
H I    & 21 cm  & 18.36 $\pm$ 0.03   & 18.43 $\pm$ 0.03     & \multicolumn{1}{c}{$<$17.26} \\
O I    & 1302   &13.99 $\pm$ 0.19    & \multicolumn{1}{c}{$<$14.27}& \multicolumn{1}{c}{$<$14.10} \\
C II   & 1334   & 14.35 $\pm$ 0.04   & \multicolumn{1}{c}{$^b$}    & 14.12 $\pm$ 0.07            \\
C IV   & 1548   & 13.85 $\pm$ 0.06   & 13.85 $\pm$ 0.06            & 13.76 $\pm$ 0.09            \\
C IV   & 1550   & 13.89 $\pm$ 0.11   & 13.75 $\pm$ 0.13            & 13.78 $\pm$ 0.15            \\
Si II  & 1190   & 13.56 $\pm$ 0.10   & \multicolumn{1}{c}{$^b$}    & 13.27 $\pm$ 0.19            \\
Si II  & 1193   & 13.44 $\pm$ 0.07   & 13.12 $\pm$ 0.12            & 13.12 $\pm$ 0.15            \\
Si II  & 1526   & 13.63 $\pm$ 0.14   & 13.29 $\pm$ 0.23            & 13.16 $\pm$ 0.35            \\
Si III & 1206   & 13.43 $\pm$ 0.03   & 13.33 $\pm$ 0.03            & 13.36 $\pm$ 0.04            \\
Si IV  & 1393   & 13.20 $\pm$ 0.15   & 13.08 $\pm$ 0.12            & 13.02 $\pm$ 0.17             \\
Si IV  & 1402   & 13.09 $\pm$ 0.24   & 13.08 $\pm$ 0.26            & 13.25 $\pm$ 0.20            \\
Al II  & 1670   & 12.44 $\pm$ 0.14   & 12.80 $\pm$ 0.10            & \multicolumn{1}{c}{$<$11.97}   \\
Fe II  & 1608   & \multicolumn{1}{c}{$<$13.48}  & \multicolumn{1}{c}{$<$13.69}& \multicolumn{1}{c}{$<$13.59} \\
S II   & 1250   & \multicolumn{1}{c}{$<14.28$}  & \multicolumn{1}{c}{$^b$}    & \multicolumn{1}{c}{$<$14.51} \\
S II   & 1253   & 14.07 $\pm$ 0.48   & \multicolumn{1}{c}{$^b$}    & 14.26 $\pm$ 0.38            \\
S II   & 1259   & \multicolumn{1}{c}{$<$13.82}    & \multicolumn{1}{c}{$<$14.13}    &  14.74 $\pm$ 0.12      \\ \hline
\end{tabular}
\begin{tablenotes}
\small
\item \textbf{Notes}
\item The uncertainty on N(H I 21 cm) is statistical. Errors on the UV metal-line column densities include both statistical 
and systematic (continuum-placement) uncertainties. Upper limits are 3$\sigma$. 
%\item $^a$ This upper limit is 5$\sigma$
\item $^b$ indicates blend.
\item $^c$ Column density measurements are made in the velocity range ($v_{min}$ to $v_{max}$) in units of km s$^{-1}$; ($-$270, $-$80) for B0117-2837, ($-$250, $-$110) for B0120-28, and ($-$280, $-$110) for Ton S210.
%\item $...$ indicates blending \& saturation
\end{tablenotes}
\label{tab N measurement}
\end{threeparttable}
\end{table}

\subsection{Column Density Measurements}
\indent The column densities for each metal line of interest were derived by measuring the absorption in the LSR velocity range of the CHVC, using the apparent optical depth (AOD) method \citep{SavSem91, SemSav92, Jen96}.  The velocity limits, $v_{min}$ and $v_{max}$, are determined by visual inspection of the data, finding the velocities where the flux recovers to the continuum on either side of the CHVC absorption component. The AOD in each pixel is calculated as   $\tau_{a}(v)=\ln[F_{c}(v)/F(v)]$ where $F_{c}(v)$ and $F(v)$ are the best-fit continuum flux and the observed flux respectively. The total AOD is calculated from the equation  $\tau_{a}=\int_{v_{min}}^{v_{max}} \tau_{a}(v)\mathrm{d}v$ and the apparent column density from $N_a(v)=3.768 \times 10^{14}(f\lambda)^{-1}\tau_{a}(v)$ where $f$ is the oscillator strength of the transition, $\lambda$ is the wavelength in \AA , and $N_a(v)$ is in units of cm$^{-2}$ (km s$^{-1}$)$^{-1}$. This method is reliable if the lines are resolved and unsaturated. It requires that two or more lines have sufficiently different $f\lambda$ to take into account the effects of saturation \citep{SavSem91}. 

\indent For non-detections, i.e. cases where there is no absorption present at $3\sigma$ significance in the chosen velocity range, we give a $3\sigma$ limit on the column density, assuming a linear curve of growth. Table \ref{tab N measurement} shows the measured CHVC column densities toward the three sightlines. Errors on column densities include both statistical uncertainty as well as continuum fitting uncertainty of 5\% as we find S/N of the data $\sim$ 20 (Table \ref{tab QSO}).

\indent Assuming that the H I 21 cm line is optically thin, its column density is calculated using $ N$(H I) $=1.823 \times 10^{18}\int_{v_{min}}^{v_{max}} T_{B} \mathrm{d}v$ \citep[e.g.][]{DiLo90}, where $N$(H I) is in units of cm$^{-2}$, T$_{B}$ is the brightness temperature in kelvin, and $v_{min}$ and $v_{max}$ are the velocity integration limits in km s$^{-1}$. We fit Gaussian components  to derive the H I column densities for each of the three directions, which match well with the results obtained from the integration method using the above-mentioned formula of \citet{DiLo90}.  Details are given in Table \ref{tab H I}. 

\begin{table}[hbt]
\resizebox{1.05\columnwidth}{!}{%
\begin{threeparttable}
\caption{CHVC H I Parameters}
\centering
\begin{tabular}{c c c c c}
\hline\hline

Parameter &   Ton S210 & B0120-28 & B0117-2837 &\\ [0.5ex]
\hline
A (K) & 0.0347 $\pm$ 0.0016 & 0.0489 $\pm$ 0.0021 & ...\\
$\sigma_{v}$ (km s$^{-1}$) & 14.64 $\pm$ 0.78 & 12.24 $\pm$ 0.61 & ...\\
$v_{0}$ (km s$^{-1})$ & $-$199.03 $\pm$ 0.78 & $-$177.98 $\pm$ 0.61 & ... \\
%FWHM (km s$^{-1}$) & 34.40$\pm$1.83 & 28.77$\pm$1.43 \\
N(H I) (cm$^{-2}$) & (2.32 $\pm$ 0.16)$\times 10^{18}$ & (2.74 $\pm$ 0.18)$\times 10^{18}$ & $<$3$\times 10^{17}$\\[1ex]
    
\hline
\end{tabular}
\begin{tablenotes}
\small
\item \textbf{Notes}
\item A : amplitude of the fitted Gaussian in K (peak H I brightness temperature), $\sigma_{v}$ : dispersion in km s$^{-1}$
\item $v_{0}$ : LSR velocity centroid in km s$^{-1}$
%\item FWHM: FWHM line width in km s$^{-1}$
\item N(H I) : H I column density in cm$^{-2}$
\item Upper limit on N(H I) is 5$\sigma$
\end{tablenotes}
\label{tab H I}
\end{threeparttable}%
}
\end{table}

\subsubsection{\emph{Ton S210}}
 
 \indent The absorption lines analyzed toward Ton S210 are listed in Table \ref{tab N measurement}. The chosen LSR velocity limits are $v_{min}$ = $-$270 km s$^{-1}$ and $v_{max}$ = $-$110 km s$^{-1}$. The lines which are well-resolved and unsaturated are C II 1334, C IV 1548, 1550, Si II 1190, 1193, 1526, Si III 1206, Si IV 1393, 1402 and Al II 1670, and so form the main part of our analysis. Their column densities are calculated using the AOD technique. The S II lines show no detection. For the abundance determination, we report the most constraining value, i.e. log [$N$(S II 1259)/cm$^{-2}$]$<$13.82, obtained by adjusting the velocity integration limits to $-$300 km s$^{-1}$ and $-$90 km s$^{-1}$. Ton S210 is the only sightline where O I 1302 is detected (Figure \ref{fig nightonly}), with a column density of log [$N_a$(O I 1302)/cm$^{-2}$] = 13.99 $\pm$ 0.19. We obtain log [N(H I)/cm$^{-2}$] = 18.36 $\pm$ 0.03 from the Parkes data.

\subsubsection{\emph{B0120-28}}

  \indent The absorption lines analyzed toward B0120-28 are listed in Table \ref{tab N measurement}. The velocity limits for the AOD integration are taken as  $-$250 km s$^{-1}$ to $-$110 km s$^{-1}$. There is no detection of O I 1302 (Figure \ref{fig nightonly}). The lines which are well-resolved and unsaturated are C IV 1548, 1550, Si II 1193, 1526, Si IV 1393, 1402.  Figure \ref{fig B0120-28} shows an absorption feature (blend) on the lower velocity end of Al II 1670 and the higher velocity end of Si IV 1402. We therefore adjust the lower velocity limit for Al II 1670 to $-$300 km s$^{-1}$  and the upper velocity limit for Si IV 1402 to $-$70 km s$^{-1}$ to obtain the column densities of these absorption lines. C II 1334, Si II 1190, S II 1250, 1253 are blended, and so are neglected. Fe II 1608 and S II 1259 show very weak absorption. The H I column density derived from the Parkes data is log [N(H I)/cm$^{-2}$] = 18.43 $\pm$ 0.03. 

\subsubsection{\emph{B0117-2837}}
\indent The absorption lines analyzed toward B0117-2837 are listed in Table \ref{tab N measurement}. The velocity limits are chosen as  $-$270 km s$^{-1}$ and $-$80 km s$^{-1}$.  O I 1302 shows no detection (Figure \ref{fig nightonly}). The lines which are well-resolved and unsaturated are C II 1334, C IV 1548, 1550, Si II 1190, 1193, 1526, Si III 1206, Si IV 1393, 1402 and S II 1259. Al II 1670 and S II 1253 show weak absorption. Fe II 1608 and S II 1250 show no absorption. We give an upper limit of H I 21 cm column density assuming that the undetected line is intrinsically of Gaussian shape, as shown in Figure \ref{fig H I}. The rms brightness temperature (sensitivity) is 7.8 mK at the original velocity resolution of about 0.4 km s$^{-1}$. Assuming a line width of 35 km s$^{-1}$ FWHM, similar to what is observed toward Ton S210, this translates into a $5\sigma$ H I column density limit of $N$(H I)$<3\times$10$^{17}$ cm$^{-2}$.
  
  \indent Among the three sight lines, Ton S210 shows the highest UV metal-line column densities.
  
  \subsection{Chemical Abundances}

\indent The measured ion column densities encode important information about the CHVC's physical and chemical conditions. We use the standard notation to describe the relative gas-phase abundance of an element $X$ with respect to neutral hydrogen, given by $[X/H]\equiv \log(X/H)_{obs}-\log(X/H)_{\odot}$
where $(X/H)_{obs}$ is the relative abundance of X with respect to H in the considered medium while $(X/H)_{\odot}$ is the relative abundance in the sun.  Neutral oxygen and neutral hydrogen have similar ionization potentials, and there is a strong charge-exchange reaction that couples both ions in gas with sufficiently high density \citep{F71}. Oxygen is also found to be relatively undepleted onto dust grains in interstellar environments \citep{Je05, Me98}. This makes O I the best ion to study the overall metallicity of neutral and weakly ionized gas. In our case, we find a significant detection of O I 1302 only toward Ton S210, which gives us [O/H] = [O I/H I] = $-$1.06 $\pm$ 0.22 (before ionization correction). We include a systematic error of 0.10 dex \citep{Fox10} on the metallicity measurements accounting for the effect of the beam-size mismatch between UV data (infinitesimal beam) and the radio data (finite beam). The other two sightlines show no detection of O I. Like O I/H I, the S II/H I ratio is also considered to be one of the most reliable among the available abundance indicators, since both of them are least affected by dust and ionization effects  \citep[although see][]{Je09}. S II absorption is not detected at significant levels in any of the three sightlines under consideration. We find [S II/H I] $<$ 0.34 toward Ton S210, which is consistent with [O/H] measurement. 

\begin{figure}[hbt]\centering
\includegraphics[trim = 0mm 90mm 0mm 90mm,clip, width=1.2\linewidth]{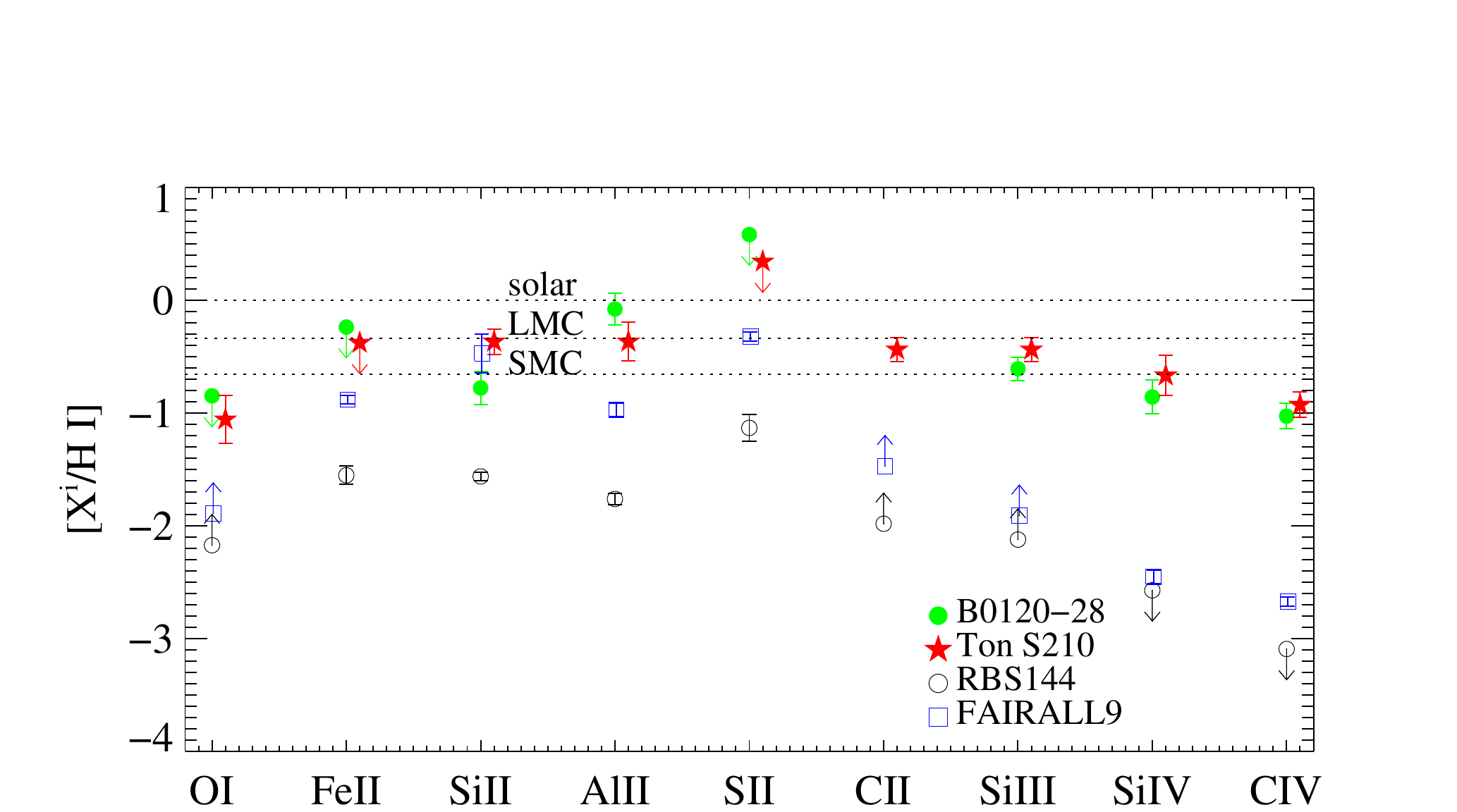}
\caption{Ion ratios for the two sightlines B0120-28 and Ton S210, and comparison with other previously studied sightlines RBS144 and Fairall 9. The solar, average LMC (0.46 solar), and average SMC (0.22 solar) abundances are plotted as horizontal lines. We note that H I column density toward these two sightlines (log [N(H I)/cm$^{-2}$]: 20.17 for RBS144 \citep{Fox13} and 19.97 for Fairall 9 \citep{Rich13}) are much higher than the H I column density toward CHVC sightlines (log [N(H I)/cm$^{-2}$]: 18.36 for Ton S210 and 18.43 for B0120-28).}
\label{fig ratio}
\end{figure}

 \indent The ion abundances of the absorbing species,  $[X/H]$ form the empirical indicators of the gas-phase abundances.  Figure \ref{fig ratio} compares the ion abundances measured toward B0120-28 and Ton S210 with the compilation of LMC and SMC reference abundances presented by \citet{RuDo92}. The solar abundance is shown by a horizontal line at [X/H] = 0. The average LMC abundance is 0.46 solar, and the average SMC abundance is 0.22 solar. They are shown as the horizontal lines. On the same plot, the MS ion abundances measured toward two other previously studied sightlines, RBS144 \citep{Fox13}, and Fairall 9 \citep{Rich13}, are also shown. 

\subsection{Ionization Modeling}
 \indent To derive the elemental abundances from the ion abundances, ionization corrections must be applied. The ionization corrections are defined as the difference between the intrinsic elemental abundance and the measured ion abundance, i.e. IC(X$^{i})$ = [X/H] $-$ [X$^{i}$/H I]. These corrections show the amount which must be added to the observed ion-to-H I ratio to determine the intrinsic abundances. A set of photoionization models using $Cloudy$ \citep[v13.02][]{Fe13} was run to investigate the process of photoionization of the low-ionization species by the incident radiation field, which helped us derive the ionization corrections.

\indent When running the $Cloudy$ model, we assume that our CHVC is a plane-parallel uniform density slab exposed to a radiation field which includes both quasars and galaxies, and use the Haardt \& Madau (2005) extragalactic background built into $Cloudy$ at a redshift of zero. Models including the contribution from ionizing photons from the Milky Way and the Magellanic Clouds are presented in \cite{Fox14}. An overall metallicity of [X/H] = $-$1 is used, (as justified by the measured [O I/ H I] ratio toward Ton S210) for each of the sightlines while the column density of the neutral gas (log [N(H I)]/cm$^{-2}$) is taken as 18.36 and 18.43 for Ton S210 and B0120-28 respectively; this defines the stopping condition of the model. The logarithmic gas density is varied  from $-$1.0 to $-$4.0 with an interval of 0.5.  This is equivalent to varying the ionization parameter log $U$ $\equiv$ log $(n_{\gamma}/n_{H}$), from $-$5.0 to $-$2.0. We use the minimum chi-square method of model-fitting to find the model (i.e. the value of log $U$) for each sightline that best reproduces the observations. No $Cloudy$ model could be made for the B0117-2837 direction because there was no significant detection of H I 21 cm emission in the CHVC toward this sightline.

\begin{figure}
\centering
\begin{subfigure}%{.5\textwidth}
  \centering
  \includegraphics[trim = 0mm 50mm 0mm 58mm,clip, width=1.3\linewidth]{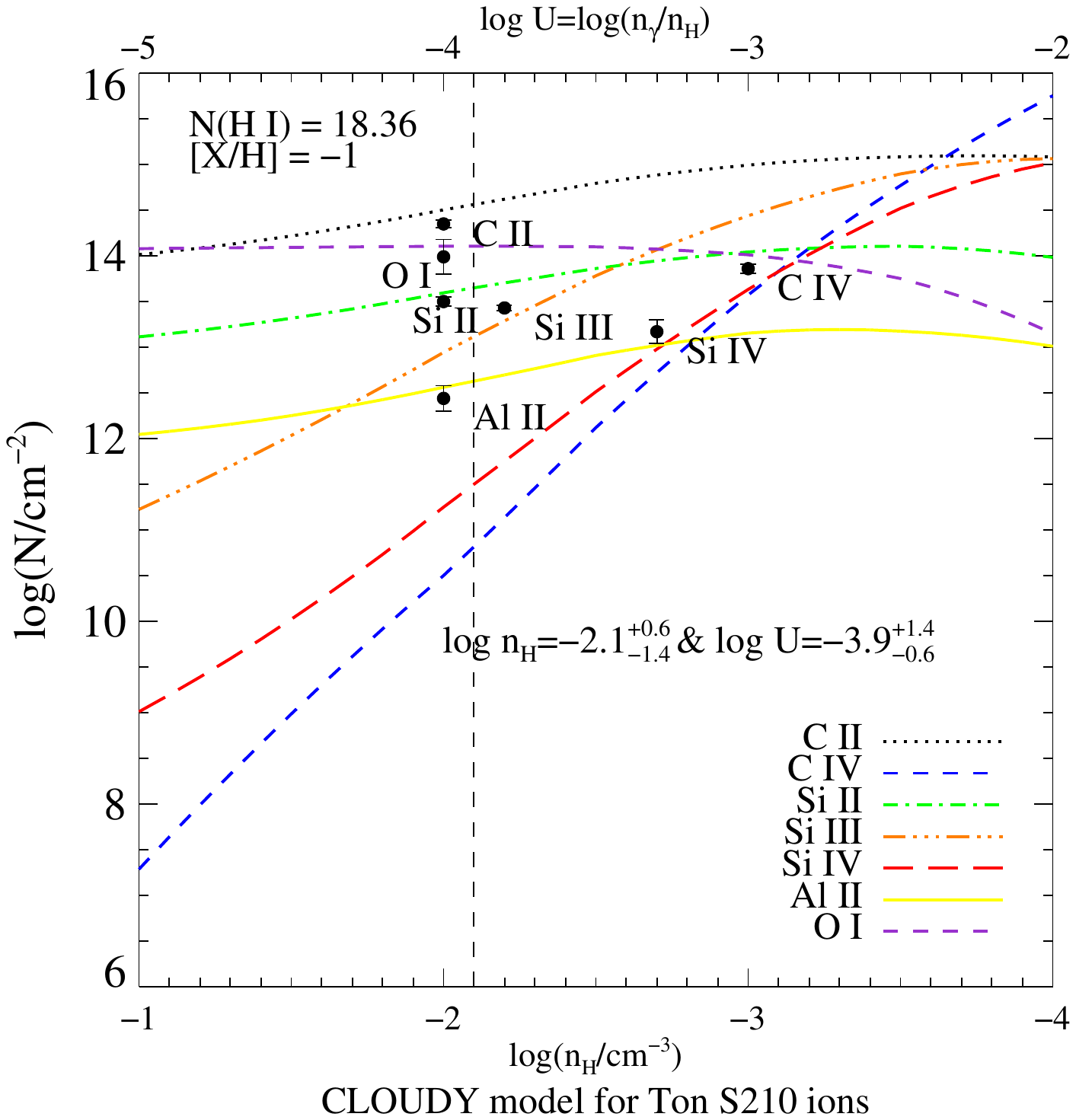}
  \end{subfigure}%
\begin{subfigure}%{.5\textwidth}
  \centering
  \includegraphics[trim = 0mm 50mm 0mm 58mm,clip, width=1.3\linewidth]{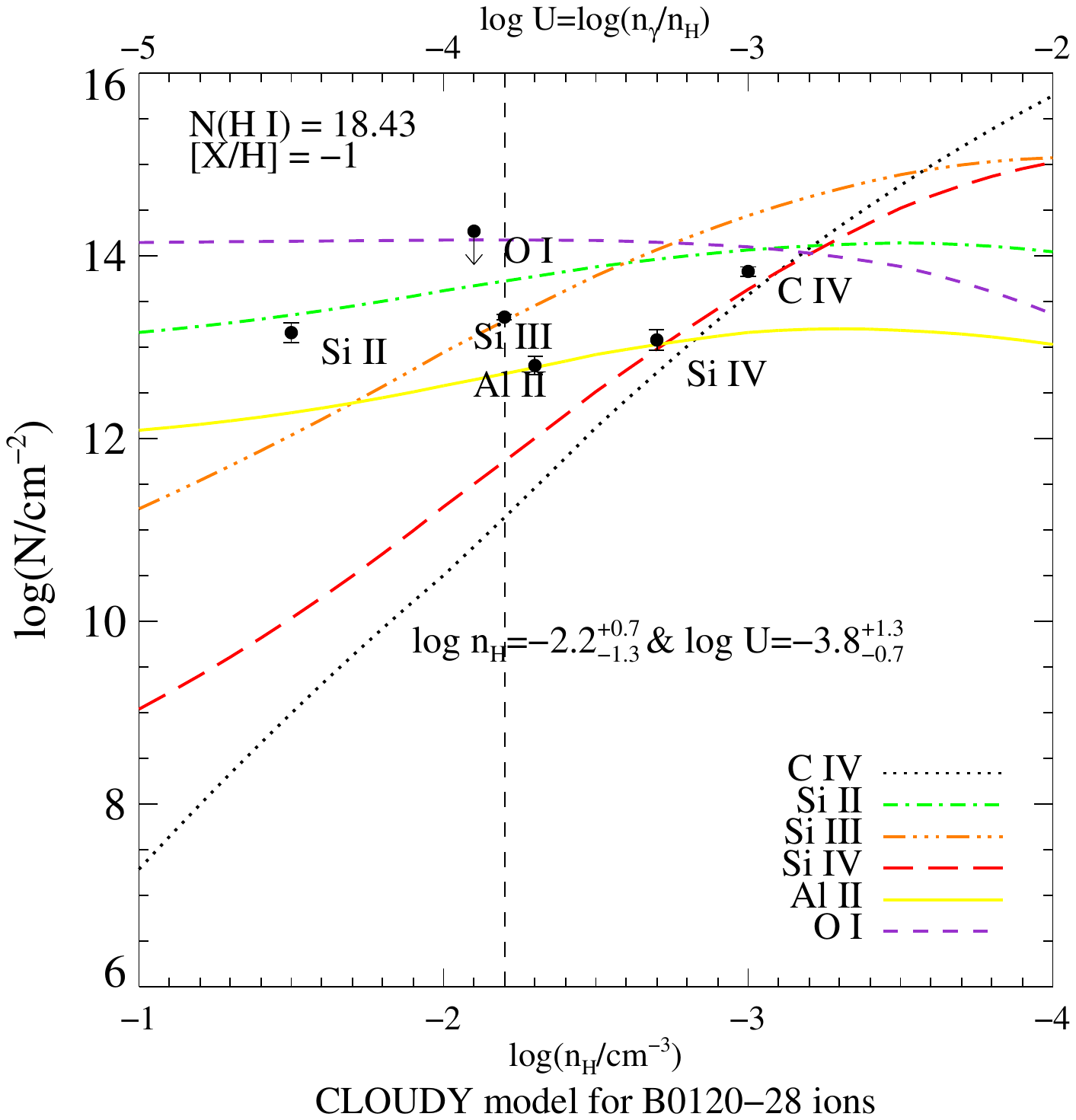}
 \end{subfigure}
\caption{$Cloudy$ photoionization models of the absorbing species observed in the CHVC toward Ton S210 and B0120-28. These models are generated assuming that the low-ionization species (C II, Si II, Al II) are photo-ionized by the incident radiation field which includes both quasars and galaxies, and the overall metallicity = [X/H] = $-$1. The colored lines show the predicted column densities of each ion while circles show the observed column densities for these ions. No $Cloudy$ model could be made for B0117-2837 because there was no significant detection of H I 21 cm emission in the CHVC in this sightline.}
\label{fig Cloudy}
\end{figure}

\begin{figure}[hbt]
\centering
\begin{subfigure}%{.5\textwidth}
  \centering
  \includegraphics[trim = 0mm 80mm 0mm 88mm,width=1.2\linewidth]{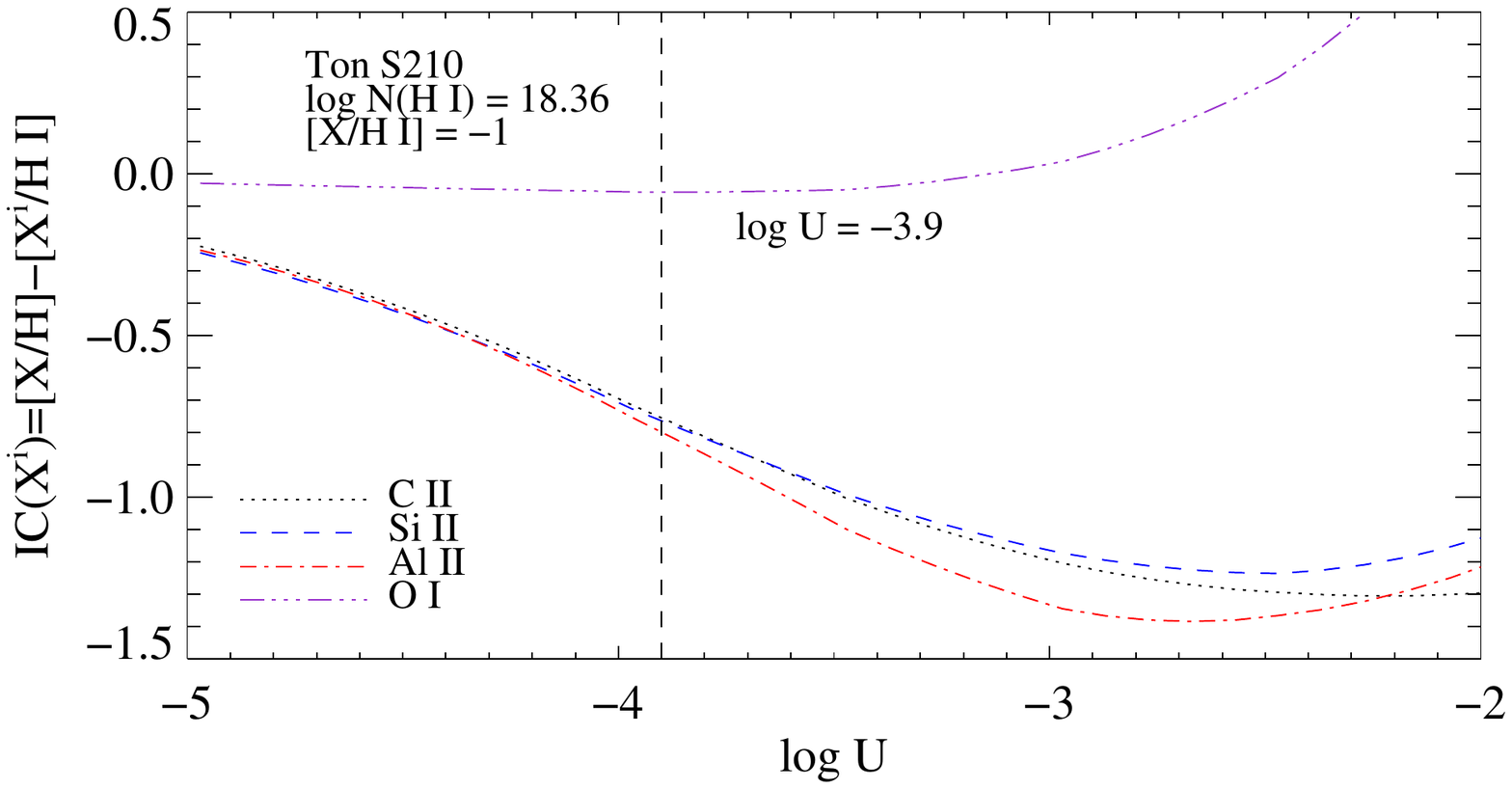}
 \end{subfigure}%
  \begin{subfigure}%{.5\textwidth}
  \centering
  \includegraphics[trim = 0mm 80mm 0mm 88mm,width=1.2\linewidth]{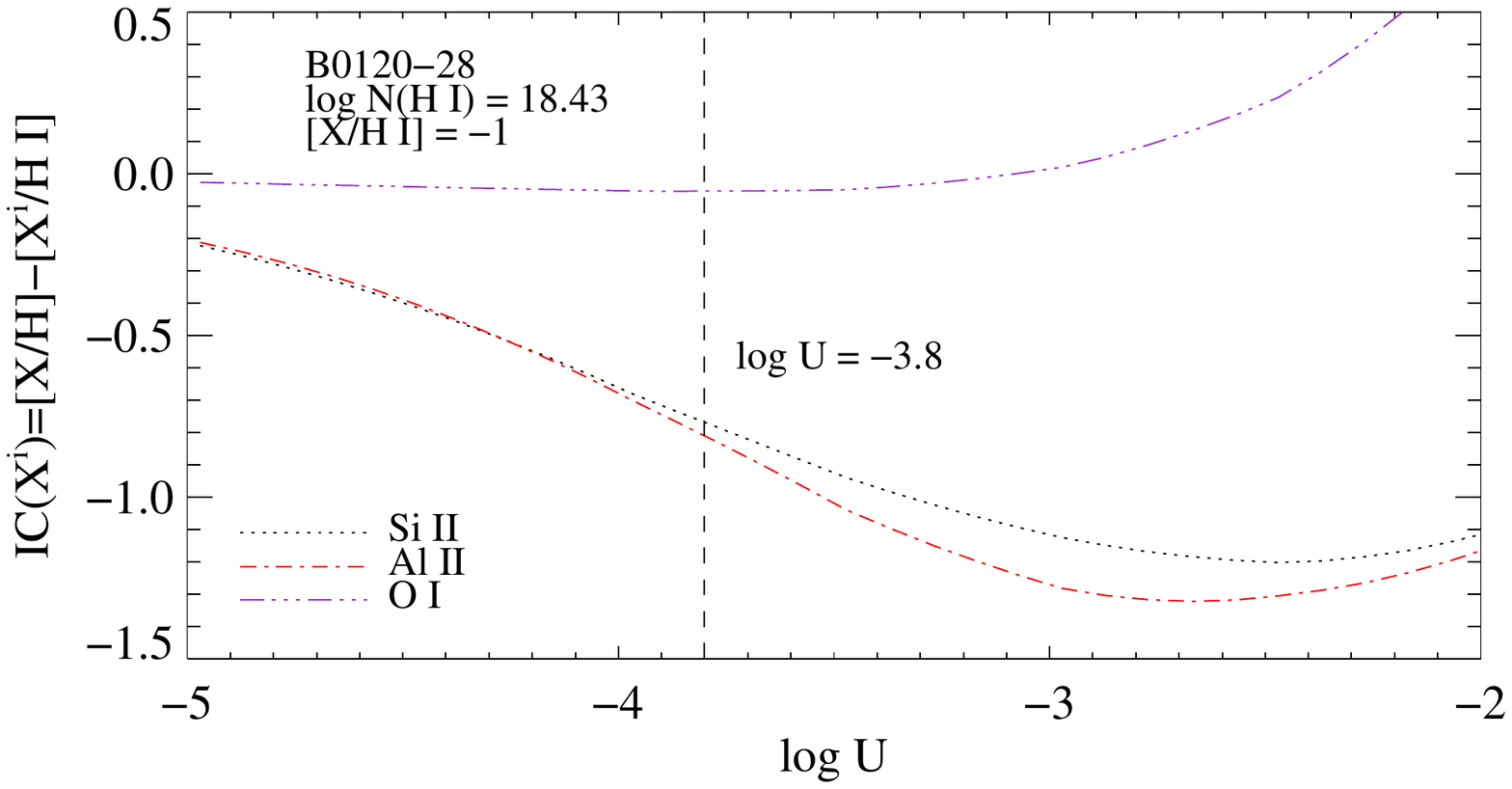}
  \end{subfigure}
\caption{Ionization corrections for low ions (C II, Si II, Al II) and O I in our $Cloudy$ models of the CHVC toward Ton S210 (top) and B0120-28 (bottom). No plot of ionization correction for C II is shown toward B0120-28 since C II 1334 is blended toward this sightline. These corrections are used to derive the elemental abundance from the ion abundance. The small ionization correction for O I  justifies its use as a robust metallicity indicator.}
\label{fig test}
\end{figure}

\indent We present the results obtained from $Cloudy$ models in two sets of figures (\ref{fig Cloudy} and \ref{fig test}), each consisting of the analysis for the two sightlines, Ton S210 and B0120-28. The first set (Figure \ref{fig Cloudy}) shows the variation of column densities of the ions with respect to the ionization parameter log $U$ and gas density log $n_H$. On the same plots, we show the observed column densities calculated from the AOD method. The best-fit values of log $U$ and of log $(n_H$/cm$^{-3}$), shown on these plots are calculated considering that only low-ionization species are photoionized by the radiation field, and not the highly ionized species (C IV and Si IV), which may have a contribution from collisional ionization \cite[e.g. ][]{Fox10}. In the text which follows, we use the term low ions to refer to singly- and doubly-ionized species (such as Si II and Si III),
and the term high ions to refer to triply-ionized species (C IV and Si IV). The second set (Figure \ref{fig test}) shows the  variation of ionization corrections for the low ions calculated from the $Cloudy$ models, as a function of the ionization parameter. We now discuss the results drawn from these two sets of figures.

\subsubsection{\emph{Cloudy Results toward Ton S210}}
\indent The photoionization model obtained from $Cloudy$ yields an ionization parameter log $U$ = $-3.9^{+1.4}_{-0.6}$, corresponding to a gas density log ($n_{H}$/cm$^{-3}$) = $-2.1^{+0.6}_{-1.4}$. The ionization corrections for C II, Si II, and Al II derived from our best-fit model are: IC(C II) = $-$0.77 dex, IC(Si II) = $-$0.78 dex, IC(Al II) = $-$0.82 dex, which are all larger (in an absolute sense) than IC(O I) = $-$0.06 dex. The low ionization correction obtained for O confirms that it is a robust metallicity indicator. Applying these ionization corrections, we derive gas-phase abundances, [C/H] = $-$1.21 $\pm$ 0.11, [Si/H] = $-$1.15 $\pm$ 0.11, [Al/H] = $-$1.19 $\pm$ 0.17, and [O/H] = $-$1.12 $\pm$ 0.22. We find that the gas-phase abundances for all four of the low ions agree within 0.09 dex. 

\subsubsection{\emph{Cloudy Results toward B0120-28}}

\indent The photoionization model obtained from $Cloudy$ yields the ionization parameter log $U$ = $-3.8^{+1.3}_{-0.7}$, corresponding to a gas density log ($n_{H}$/cm$^{-3}$) = $-2.2^{+0.7}_{-1.3}$. As we found for Ton S210, the ionization corrections toward B0120-28 for Si II, and Al II derived from our best-fit model are significant: IC(Si II) = $-$0.78 dex, IC(Al II) = $-$0.83 dex. Applying these ionization corrections, we derive gas-phase abundances,  [Si/H] = $-$1.56 $\pm$ 0.15, and [Al/H] = $-$0.91 $\pm$ 0.14. We find IC(O I) = $-$0.05 dex, which gives [O/H] $<$ $-$0.90. We do not report [C/H] as C II 1334 is blended toward B0120-28. 

\indent Figure \ref{fig Cloudy} also shows that the best-fit photoionization $Cloudy$ model can only explain the low-ion column densities, but not the high-ion column densities-the C IV and Si IV are not explained near a gas density of log $n_H$ = $-$2.1 (Ton S210) and $-$2.2 (B0120-28), since the models at that density predict much smaller column densities of these two ions than are observed. This result has been found previously in the context of other HVCs by many authors \citep{Col05, Fox05}. This shows the need of a separate ionization mechanism (possibly collisional ionization) to explain the high ions. The difference in velocity structure of the low and high ions also indicates the separate behaviour of high ions. 
   
\begin{figure}[hbt]\centering
\includegraphics[trim = 0mm 90mm 0mm 90mm,clip, width=1.2\linewidth]{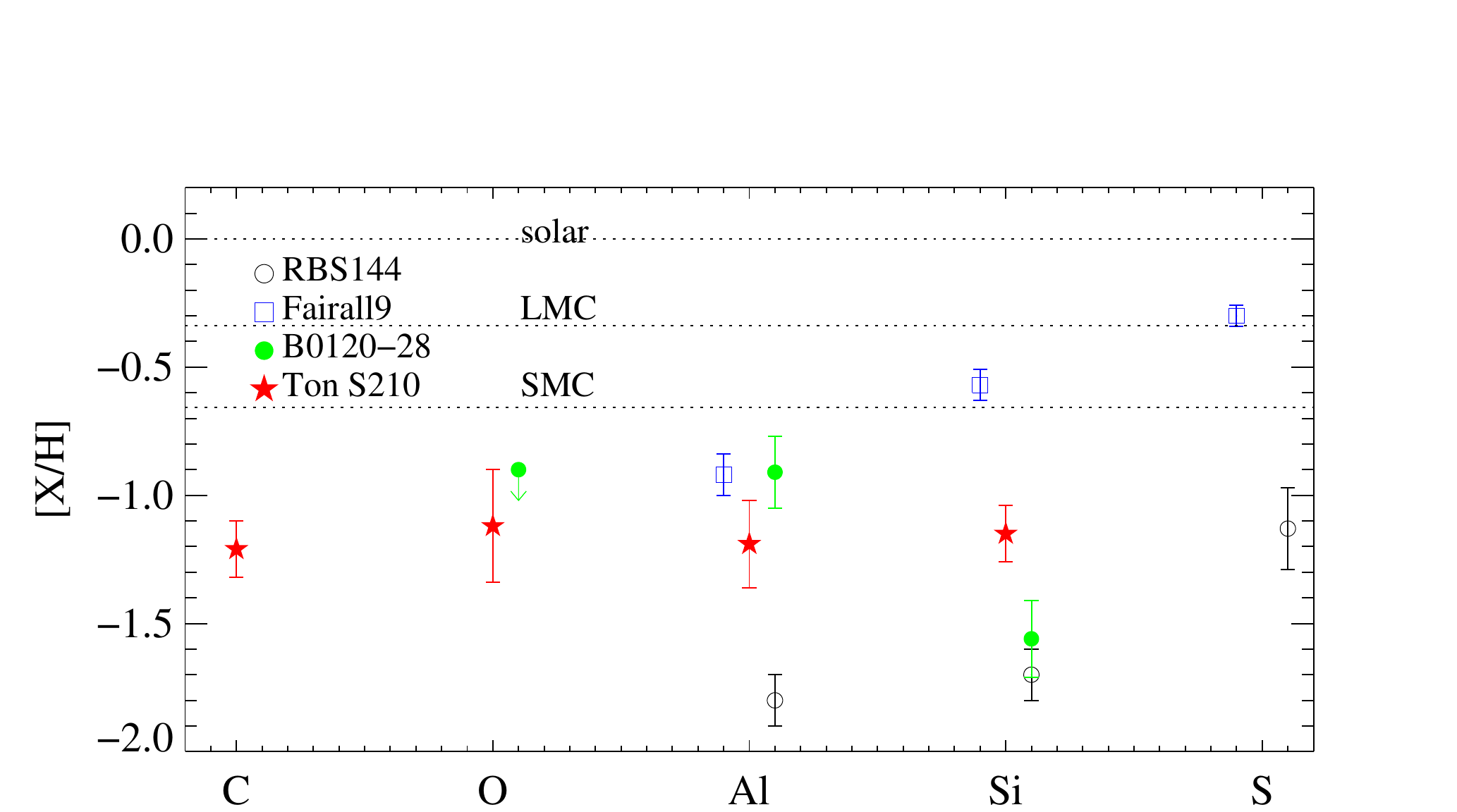}
\caption{Gas-phase abundance ratios for the two sightlines B0120-28 and Ton S210, and comparison with other previously studied sightlines RBS144 and Fairall 9. The solar, average LMC (0.46 solar), and average SMC (0.22 solar) abundances are plotted as horizontal lines.}
\label{fig element}
\end{figure} 

\section{Discussion}
\indent Our study is designed to investigate if the CHVC  under study lying 10$^{\circ}$ off the main body of the Magellanic Stream is physically connected with it and to study the ionization conditions across the cloud. We compare our findings to those obtained for RBS144 and Fairall 9 which lie behind the main body of the Magellanic Stream but those directions have much higher H I column densities (log [N(H I)/cm$^{-2}$]: 20.17 for RBS144 \citep{Fox13} and 19.97 for Fairall 9 \citep{Rich13}). From Figure \ref{fig ratio}, we find that the ion abundances derived for our sightlines, B0120-28 and Ton S210 are above  those obtained for RBS144 and Fairall 9. To explore this further, we solved for the ionization correction of the low ions using $Cloudy$ photoionization models with  one tenth solar metallicity in the CHVC, as measured by O I/H I toward Ton S210, thus deriving the elemental abundances of these species. Figure \ref{fig element} shows gas-phase abundance ratios for the two sightlines B0120-28 and Ton S210, and comparison with other previously studied sightlines RBS144 and Fairall 9. The derived O abundance after correction, [O/H]$_{\mathrm {CHVC}}$ = $-$1.12 $\pm$ 0.22, is close to the metallicity of the same CHVC reported by \citet{Rich09} using HST/STIS data, [O/H] = $-$1.01. The metallicity measurements of the Magellanic Stream from previous studies give $-$1.13 $\pm$ 0.16  for RBS144 \citep{Fox13},  $-$1.24 $\pm$ 0.12  for NGC 7714 and $-$1.00 $\pm$ 0.09 for NGC 7469 \citep{Fox10}, all close to 0.1 solar as in our case, while the metallicity measurement toward Fairall 9 is a factor of five higher $-$0.30 $\pm$ 0.04 \citep{Rich13, Gi00}. It should be noted here that these three sightlines (RBS144, NGC 7714 \& NGC 7469) probe the main body of the MS, and have high H I column density (20.17 for RBS144), while our CHVC toward Ton S210 lies 10$^{\circ}$ off the MS with low H I column density (18.36). The red stars in Fig \ref{fig element} (Ton S210) form a fairly flat line near [X/H] $=$ $-$1.2, but the black circles (RBS144)
and blue squares (Fairall 9) show a depletion pattern where [Al/H] $<$ [S/H] and [Si/H] $<$ [S/H], indicating
that Al and Si are depleted relative to S in those directions. This apparent difference in dust depletion
is likely related to the much higher N(H I) for the RBS144 and Fairall 9 directions compared to the Ton S210 direction.

\indent The fraction of silicon in each ionization state can be calculated for each sightline. For B0120-28 and B0117-2837, the ratios of Si II, Si III and Si IV with respect to the total Si are found to be  0.30, 0.45 and 0.25 respectively. For Ton S210, the ratios of Si II, Si III and Si IV with respect to the total Si are found to be about 0.43, 0.36 and 0.20 respectively. This shows that there is not much variation in the ionization breakdown of silicon from one part of the cloud to the other. 

\indent Our photoionization models underestimate the observed column densities of the high ions C IV and Si IV, suggesting that these ions live in a separate phase of gas than the low ions. Strikingly, we find that the observed
values of log C IV/Si IV toward B0117-2837, B0120-28 and Ton S210 are 0.65 $\pm$ 0.15, 0.75 $\pm$ 0.12 and 0.69 $\pm$ 0.14 respectively,  which are very close to the ratios measured in the main body of the Magellanic Stream, 0.69 $\pm$ 0.04 toward NGC 7469 and 0.77 $\pm$ 0.13 toward Mrk 335 \citep{ 
Fox10}. This similarity suggests that the mechanism involved in producing the high ions in the MS, most probably collisional ionization, is at work in the CHVC under study too.

\indent Our finding that the CHVC metal abundance of 0.1 solar matches the abundance measured along the main body of the Stream suggests that the CHVC has its origin in the Stream. This is reinforced by the similar central velocity of the CHVC and the Stream in this part of the sky. The CHVC may represent a fragment of the Stream that has been removed by its interaction with the surrounding plasma. Such interaction is invoked to explain the small-scale structure seen in high-resolution 21 cm maps of the Stream \citep{WeKo08, St02, St08}, and is predicted by hydrodynamical simulations \citep{Bh07, HePu09}.

\section{Summary}

In this paper, we have analyzed UV absorption-line data from HST/COS and radio H I 21 cm emission data from the Parkes telescope in three directions (Ton S210, B0120-28 and B0117-2837) passing through CHVC 224.0-83.4-197. This has allowed us to study the chemical composition and the physical conditions of the CHVC, which lies near the edge of the Magellanic Stream near the South Galactic Pole. The main results obtained from the work are as follows:

\begin{enumerate}%[noitemsep]

\item For all three sightlines, absorption is detected in one individual absorption component centered at $v_{LSR}$ = $-$200 km s$^{-1}$. For B0117-2837, a second high-velocity component is seen at $-$120 km s$^{-1}$, although there is no evidence that this component is connected to the CHVC.

\item Detected atoms and ions in the CHVC include C II, C IV, Si II, Si III, Si IV and Al II. Absorption in O I 1302, a good metallicity indicator, is observed using a night-only reduction of the $COS$ data of Ton S210, and can be detected only toward  Ton S210. S II, Fe II, N I and N V are not detected at reliable confidence levels.

\item The H I column density in the CHVC is not high enough for significant detections in the LAB \citep{Ka05} and GASS \citep{Mc09} Surveys. Deep 21-cm Parkes observations provide the H I column density, log [N(H I)/cm$^{-2}$] as 18.43 $\pm$ 0.03 for B0120-28 and 18.36 $\pm$ 0.03  for Ton S210. For B0117-2837, an upper limit $<$17.26 corresponding to the $3\sigma$ value is reported. 

\item We calculate the ratios of Si II, Si III, Si IV to the total Si for the three sightlines under study. N (Si II)/ N (Si) varies from 0.43 (Ton S210) to 0.30 (B0120-28 and B0117-2837). We find N (Si III)/ N (Si) to be 0.36 for Ton S210, and 0.45  for B0120-28 and B0117-2837. N (Si IV)/ N(Si) varies from 0.20 (Ton S210) to 0.25 (B0120-28 and B0117-2837). Thus the ratios vary by 5-13\%, which suggests that there is not much variation of the ionization structure from one part of the cloud to the other. 

\item Using photoionization models generated from $Cloudy$ and combining the results with the observed column densities of the ten absorbing atoms and ions, the gas density is calculated. The values of the total gas density, log $(n_{H}$/cm$^{-3}$) are reported to be $-2.1^{+0.6}_{-1.4}$ and $-2.2^{+0.7}_{-1.3}$ for Ton S210 and B0120-28 respectively. Corresponding values of the ionization parameter log $U$ are $-3.9^{+1.4}_{-0.6}$ and $-3.8^{+1.3}_{-0.7}$. Using these, we calculate ionization corrections for C, Al and Si, which are significant as we expected from the low H I column density, but for O I we find IC = $-$0.06 dex. \emph{We confirm that the ionization correction is small for O I.}    

\item  Toward Ton S210, we derive gas-phase abundances, [C/H] = $-$1.21 $\pm$ 0.11, [Si/H] = $-$1.15 $\pm$ 0.11, [Al/H] = $-$1.19 $\pm$ 0.17 and [O/H] = $-$1.12 $\pm$ 0.22, which all agree to within 0.09 dex. Toward B0120-28, we derive gas-phase abundances [Si/H] = $-$1.56 $\pm$ 0.15, [Al/H] = $-$0.91 $\pm$ 0.14 and [O/H] $<-$ 0.90.

\end{enumerate}

The CHVC under study is not an isolated object, but instead belongs an extended network of filaments lying at angular distances of up to 20$^{\circ}$ from the main body of the MS \citep{WeKo08}. Our chemical abundance analysis shows that the CHVC, and by extension the population of filaments, likely originates in the Magellanic Stream. Gaseous fragments such as the CHVC may trace the disintegration and incorporation of the Stream into the Galactic halo.

\acknowledgments
\section*{Acknowledgements}
Support for program 12204 was provided by NASA through grants from the Space Telescope Science Institute, which is operated by the Association of Universities for Research in Astronomy, Inc., under NASA contract NAS~5-26555. The Parkes radio telescope is part of the Australia Telescope National Facility which is funded by the Commonwealth of Australia for operation as a
National Facility managed by CSIRO.

%-----------------BIBLIO ----------------------------  

\appendix
\renewcommand\thefigure{\thesection A.\arabic{figure}} 
\setcounter{figure}{0}

\renewcommand\thetable{\thesection A.\arabic{table}} 
\setcounter{table}{0}
%\section{Appendix}

\begin{table}[hbt]

\caption{Details of the night-only data reduction}
\centering
\begin{tabular}{@{}lllll@{}}
\hline\hline
Target            & \begin{tabular}[c]{@{}c@{}}Exposure time (s)\\               full (day+night)\end{tabular} & \begin{tabular}[c]{@{}c@{}}Exposure time (s)\\               Night-only\end{tabular}  & \begin{tabular}[c]{@{}c@{}}\% of \\               data used\end{tabular} &  \\ \hline

Ton S210          & 5047       & 2188      & 43.4\%                                                                      &  \\
B0120-28          & 5184       & 1934       & 37.4\%                                                                      &  \\
B0117-2837        & 5234       & 2090      & 39.9\%                                                                      &  \\ \hline
\end{tabular}
\label{tab O I}
\end{table}    

%channel maps
\begin{figure*}[hbt]\centering % Using \begin{figure*} makes the figure take up the entire width of the page
\includegraphics[width=\linewidth]{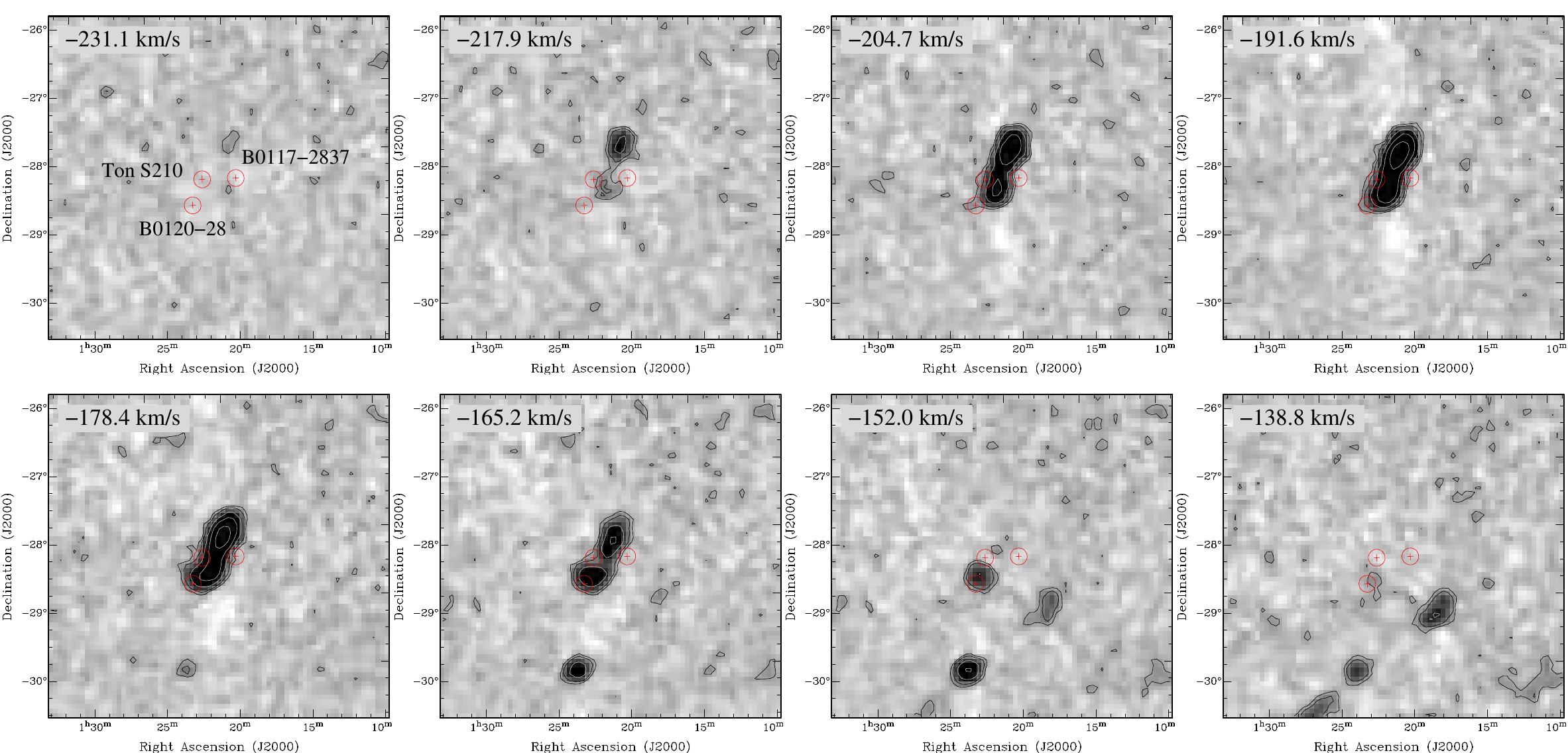}
\caption{HIPASS  channel maps of the H I 21 cm emission in the region of the sky surrounding the CHVC under study. The data shown have a velocity resolution of 26 km s$^{-1}$ and the outer contour is at a flux density level of 10 mJy. The separation in velocity between each panel is 13 km s$^{-1}$. The red circled crosses on the maps are the theee sight lines, B0117-2837, B0120-28 and Ton S210. The CHVC can be seen near the centre of the map in between $-$217.9 and $-165.2$ km s$^{-1}$. }
\label{fig channel map}
\end{figure*}

\end{document}